\documentclass[twocolumn,aps,showkeys]{revtex4}
\usepackage{times}
\usepackage{natbib}
\usepackage{graphicx}
\begin{document}
\title{Time-resolved and state-selective detection of single freely falling atoms}
\author{Torsten Bondo}
\altaffiliation[Current address: ]{Danish National Space Center, Juliane Maries Vej 30, 2100 Copenhagen, Denmark}
\author{Markus Hennrich} 
\altaffiliation[Current address: ]{Institut de Ci\`{e}nces Fot\`{o}niques, Parc Mediterrani de la Tecnologia, 08860 Castelldefels (Barcelona), Spain.}
\author{Thomas Legero}
\altaffiliation[Current address: ]{Physikalisch-Technische Bundesanstalt, Bundesallee 100, 38116 Braunschweig, Germany}
\author{Gerhard Rempe} 
\author{Axel Kuhn}
\email[Corresponding author: ]{axel.kuhn@mpq.mpg.de}
\affiliation{Max-Planck-Institut f\"{u}r Quantenoptik,\\ Hans-Kopfermann-Str. 1, 85748 Garching, Germany}
%\corauth[email]{Corresponding author. Email address: Axel.Kuhn@mpq.mpg.de}
\begin{abstract}
We report on the detection of single, slowly moving Rubidium atoms using laser-induced fluorescence. The atoms move at 3\,m/s while they are detected with a time resolution of 60\,$\mu$s. The detection scheme employs a near-resonant laser beam that drives a cycling atomic transition, and a highly efficient mirror setup to focus a large fraction of the fluorescence photons to a photomultiplier tube. It counts on average 20  photons per atom.
\end{abstract}
\date{\today}
%\PACS 07.77.Gx \sep 32.70.Jz \sep 39.30.+w
%\end{keyword}
\maketitle

\section{Introduction}
A tremendous number of experiments in atomic physics and quantum optics employ state-selective atom-detection methods. An overview of the various detection techniques can be found in  \cite{review1}. In many cases, these schemes are based on the interaction of  atoms with laser light that  excites an atomic resonance \cite{review2}. The  atoms manifest themselves by absorption, emission and dispersion of radiation. All of these effects can be observed. However, for moving atoms, the number of scattered photons  is usually  too small to discriminate the presence of a single one. The situation is different for trapped ions or atoms, where a closed transition between two energy levels can be excited. In this case, the atom cycles between two levels and spontaneously emits photons at a constant rate. This allows one to detect individual particles, provided the interaction time with the exciting laser is long enough. Such schemes have been demonstrated successfully in several experiments, where either a single or a small number of atoms was observed in a dipole trap or a magneto-optical trap \cite{motlif1,motlif2,motlif3,motlif4,motlif5}, or  a single ion in a Paul-trap \cite{ionlif1,ionlif2,ionlif3}. Even single molecules in  solvents and solids have been detected in this way, since their relaxation rate to the ground state is so fast that they can emit more than one photon while they  are observed \cite{mollif1,mollif2}. 

For freely moving atoms, detection methods are more sophisticated. For instance,  it has been shown that a single atom's dispersion is sufficient to switch the transmission of a high-finesse Fabry-Perot type optical cavity when it travels through the cavity mode \cite{Mabuchi96, Munstermann99:2,Shimizu02,Sauer04,Oettl05}. Moreover, the  Purcell-enhanced emission into a cavity has been used to trace single atoms \cite{McKeever03:2,Nussmann05,Nussmann05:2}, and also to deterministically generate single photons from a single atom located in an optical cavity \cite{Kuhn02,McKeever04,Keller04,Oxborrow05}. Although these cavity-QED based methods are very elegant and effective, they are experimentally hard to implement. Other methods to detect a moving atom  are often based on the release of a large amount of internal energy. This applies, in particular, to metastable atoms or  ions. However, the detection is destructive and does rarely allow one to determine the inner state of the particle. Only the extension to a multi-step detection method, like resonance-enhanced multi-photon ionization \cite{rempi1,rempi2},  enables one to detect individual particles in a state-selective manner. 
%Nonetheless, the ionization process is not very efficient, and only a small fraction of the formerly neutral atoms gives rise to a signal.

In this article, we focus on laser-induced fluorescence which allows us to monitor individual, freely falling atoms in a time-resolved and state-selective manner. The atoms are released from a magneto-optical trap (MOT) and  arrive at the detector in free fall with a velocity of $v=3\,$m/s. The motion of the atoms  limits the interaction time with the exciting laser beam to about 60$\,\mu$s. Within this  time interval, a significant number of photons  is recorded from every atom, so that its presence can be  discriminated from the background noise. This is achieved with a highly efficient collection system for the emitted photons.

\section{Collecting Photons}
\begin{figure}[tbh]
    \centering
    \includegraphics[width=\columnwidth]{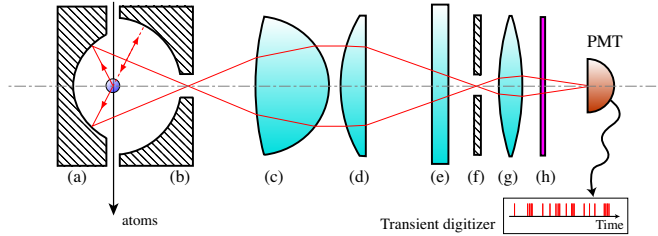}
    \caption{Fluorescence collection and imaging system (not to scale):  
    An   ellipsoidal mirror \textbf{(a)} and a spherical mirror 
    \textbf{(b)} collect the photons which are emitted from atoms that are excited in one focal 
    point of the ellipsoid. The photons are imaged 
    to the second focal point 
    of the ellipsoid. From there, a telescope consisting of lenses 
    \textbf{(c)} and \textbf{(d)} focuses them through the window of 
    the vacuum chamber \textbf{(e)}  onto an aperture \textbf{(f)}, 
    which acts as a spatial filter. Finally, they are imaged by the 
    biconvex lens \textbf{(g)} through a bandpass filter \textbf{(h)} onto the 
    cathode of a photomultiplier tube \textbf{(PMT)}. The events from 
    the PMT are recorded by a transient digitizer.}
    \label{mirror}
\end{figure}

As spontaneous emission is not directed, an efficient detector must cover a large solid angle to collect a significant amount of fluorescence photons. Our collection optics resembles the one that has been developed  for the spectroscopic detection of fast molecules \cite{Bergmann79, Shimizu83, Bergmann88}. Figure \ref{mirror} shows our setup, which is a combination of an ellipsoidal and a spherical coated aluminium mirror. It covers a solid angle close to $4\pi$, provided the atoms fluoresce at the left focal point of the ellipsoid. Photons going to the left hit the ellipsoidal mirror ($\o\,39\,$mm, eccentricity $a/b=26.4\,$mm$ / 22.7\,$mm) and are focused to the right focal point of the ellipsoid. Photons going to the right hit the spherical mirror ($\o\,50\,$mm),  are reflected back to their origin, and leave this point to the left towards the ellipsoidal mirror. Finally, they are also focused to the right focal point of the ellipsoid. However, the atoms are not well localised but move through the probe laser beam that excites them in a fluorescing volume of about 1\,mm$^3$ around the first focal point. In addition, holes in the mirrors are necessary to allow the probe beam and the atoms to enter the detector, and to finally extract the fluorescence photons. These geometrical restrictions limit the collection  efficiency to about $80\,\%$. In addition to these limitations,  the surface of the hollow mirrors in our experiment is far from being perfect. The coated aluminium surface shows a reflectivity of only 80\,\%, which in turn limits the total collection efficiency of the mirror setup to about 64\,\%.

Furthermore, the design of our apparatus did not allow placing the cathode of a photomultiplier tube (PMT) directly in the second focal point. A telescope consisting of an aspheric condenser lens ($f=35\,$mm, $\o\,50\,$mm), a planoconvex lens ($f=100\,$mm, $\o\,50\,$mm), and a biconvex lens ($f=70\,$mm, $\o\,60\,$mm) is used to image the photons onto the PMT, which is located outside the vacuum chamber. The total distance between the atoms and the PMT cathode is 50\,cm. To reduce the background of ambient visible radiation to an acceptable level, we use a 10\,nm wide band-pass filter. Additionally, a spatial filter in the  image plane of the telescope (between the second and third lens) serves to reduce light scattered from the edges of the hollow mirrors.

All the elements of the imaging system lead to a further attenuation of the light. The telescope lenses and the  window of the vacuum chamber have a transmission of 99\,\% each, the band-pass filter transmits 70\,\%, and the uncoated double-window of the PMT's cooling chamber 85\,\%. We must also take into account that the rather large observation volume is not focused perfectly onto the PMT, since it is not completely covered by the acceptance angle of the condenser lens. We estimate that as much as 1/3 of the photons are lost by these geometric effects. Hence, the collection efficiency of the whole setup (mirrors, lenses, windows and filters) amounts to  about 25\,\%. The remaining photons are finally imaged onto the photo-cathode of either a Burle PMT with a quantum efficiency of 6\,\% (for the data shown in Fig.\,\ref{bursttime}), or a Hamamatsu R943-02 PMT with a quantum efficiency of 12\,\% (all other data). Therefore the overall photo-detection efficiency is estimated to be about 1.5\,\% or 3\,\%, respectively. Note that avalanche photo diodes would provide a much higher quantum efficiency, but at the cost of a significantly smaller sensitive area. Hence we cannot use them for the detection of moving atoms. 

\begin{figure}[tbh]
    \centering
    \includegraphics[width= \columnwidth]{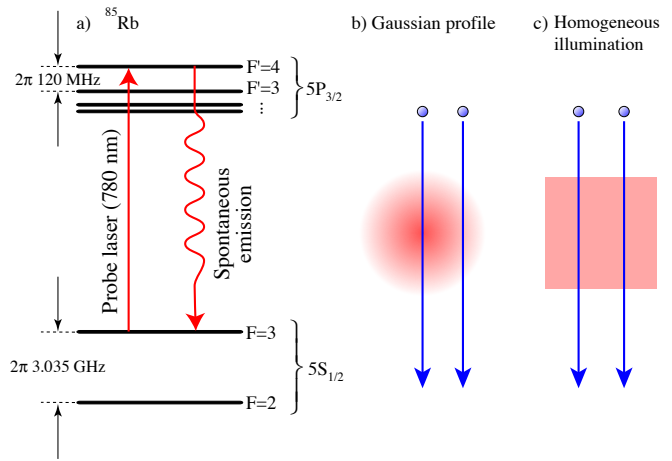}
    \caption{Detection of $^{85}Rb$ atoms by laser-induced fluorescence: 
    \textbf{(a)} The  $5S_{1/2}(F=3) \longleftrightarrow 5P_{3/2}(F'=4)$ 
    transition provides a (nearly) closed two-level system and is used to probe the atoms.
    \textbf{(b)} Atoms traveling through a Gaussian probe beam are 
    exposed to different peak intensities depending 
    on their individual trajectories. This gives rise to an ambiguous 
    fluorescence signal, since the duration of the photon burst and the 
    number of photons per atom depend on the atom's trajectory with 
    respect to the beam axis.
    \textbf{(c)} A clear signal is obtained from atoms interacting with 
    a rectangular, homogeneously illuminated area, since all atoms experience identical
    conditions. The number of emitted photons does not depend on 
    the atom's trajectory, and the interaction time is well-defined.
    To achieve these conditions, we illuminate a small rectangular 
    aperture with a large laser beam and image the aperture onto the 
    atoms.}
    \label{rb85}
\end{figure}

\section{Fluorescence Signal}
\begin{figure*}[tbh]
    \centering
    \includegraphics[width= 14cm]{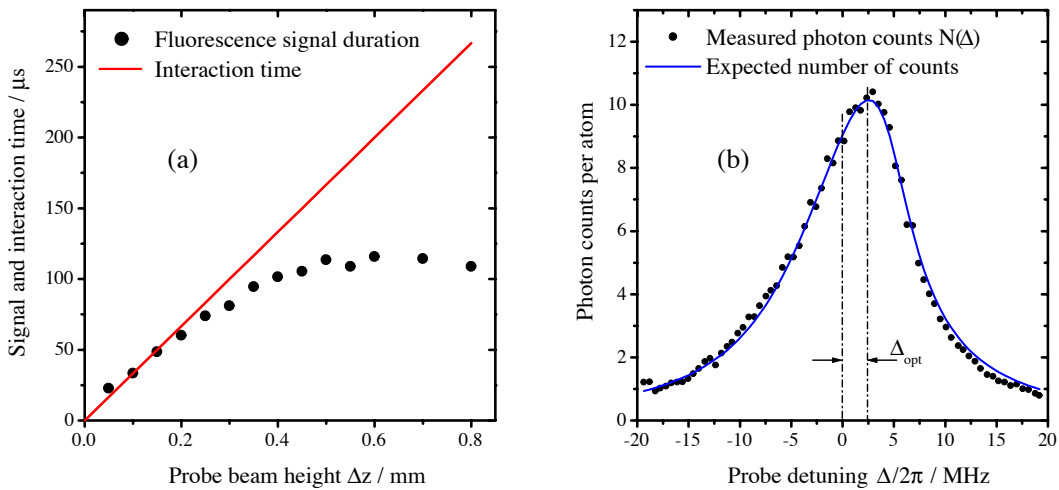}
    \caption{Fluorescence duration and spectrum for a three-fold saturated 
    probe transition: 
    \textbf{(a):} Duration of the observed fluorescence from individual 
    atoms, which are crossing the probe beam with a perpendicular velocity
    $v_{\perp}=3\,$m/s, as a function of the probe-beam height, 
    $\Delta z$, for $\Delta=+\Gamma$. 
    The solid line shows the transit time of the atoms through 
    the rectangular beam. 
    \textbf{(b):} Average number of photon counts, $N(\Delta)$, per 
    atom in $\Delta\tau=50\,\mu$s as a function of 
    the probe detuning, $\Delta$. The solid line shows the 
    expected photon number, $N_{phot}(\Delta)$, which is obtained by 
    numerically solving Eq. (\ref{Nphot}). Its amplitude is scaled down by a factor 58
    to meet the experimentally observed count rate, which is reduced due to the limited overall detection efficiency of the system.
    Note that the PMT used to measure this data had a quantum efficiency of only 6\%. All other data has been registered with another PMT that was two-times more sensitive. }    
    \label{bursttime}
    \label{detune}
\end{figure*}
Many atomic species provide a closed two-level system, i.e. a non-decaying stable energy level and an electronically excited level, which decays exclusively  to the stable one. In a dilute gas, collisional losses can be neglected, and if the transition between the two levels is driven by a probe laser, the atom cycles between these levels, and a photon is emitted spontaneously whenever a decay from the excited level occurs. In our experiment, the probe laser with a wavelength of $\lambda=780\,$nm excites the $5S_{1/2}(F=3)\longleftrightarrow 5P_{3/2}(F'=4)$ transition of $^{85}$Rb (see fig.\,\ref{rb85}). 
Thus, observation of fluorescence detects an atom in the state $5S_{1/2}(F=3)$, whereas, e.g., the second
stable state $5S_{1/2}(F=2)$ does not fluoresce.
In the limit of strong saturation and  resonant excitation, the spontaneous photon-scattering rate, $R_P$, approaches $\Gamma/2$ per atom, where $\Gamma$ is the spontaneous decay rate of the excited level. However, the probe laser intensity, $I$, is finite, the probe laser can be detuned from the atomic resonance by an amount $\Delta$, and any velocity of the atom with respect to the probe beam, $v_{||}$, gives rise to a Doppler shift. Taking power broadening into account \cite{Allen75}, the photon emission rate reads
\begin{equation}
    R_P(I,\Delta,v_{||}(t)) =     
    \frac{\Gamma}{2}\frac{I}{I+I_{sat}\left(1+\left(\frac{\Delta-k v_{||}(t)}{\Gamma/2}\right)^2\right)},
\label{Gamma}
\end{equation}
where $k=2\pi/\lambda$ and the saturation intensity, averaged over all magnetic sublevels of the considered transition, is $I_{sat}=3.9\,$mW$/$cm$^2$. We assume that $I$ and $\Delta$ do not change during the interaction of the atoms with the probe beam, but we  take into account that the atoms are accelerated because of light pressure. For each spontaneously emitted photon (with an average momentum transfer of zero), a probe photon is absorbed and the atom's velocity changes by the recoil velocity, $v_{rec}=\hbar k/m$, in the direction of the  laser beam. This acceleration continues according to
\begin{equation}
    \frac{d}{dt}v_{||}=v_{rec} R_P(I,\Delta,v_{||}(t)),
    \label{dvdt}
\end{equation}
until the atom is Doppler shifted out of resonance and stops fluorescing. Figure \ref{bursttime}(a) shows the average fluorescence duration of $^{85}$Rb atoms as a function of the probe-beam height, $\Delta z$. The interaction time of the atoms with the beam is given by $\Delta\tau=\Delta z/v$. 
For $\Delta\tau\le 60\,\mu$s, the dependence is almost linear and the fluorescence duration equals $\Delta\tau$. For longer interaction times, the fluorescence duration is limited to $120\,\mu$s, due to  the acceleration of the atoms and also possible losses to the $5S_{1/2}(F=2)$ level, as will be discussed later. Hence, we must integrate Eqs.\,(\ref{Gamma},\ref{dvdt}) in order to obtain the total number of fluorescence photons per atom, 
\begin{equation}
    N_{phot}=\int_{0}^{\Delta\tau} R_P(I,\Delta,v_{||}(t)) dt.
    \label{Nphot}
\end{equation}
Figure \ref{detune}(b) shows that the acceleration has a significant influence on the number of emitted photons, even for $\Delta\tau$ as small as $50\,\mu$s. The maximum number of photons is counted when the probe beam is  blue detuned from the atomic resonance by $\Delta_{opt}=2\pi \times 2.4\,$MHz (i.e. $\Delta_{opt}=0.4 \Gamma$, with $\Gamma (^{85}$Rb$)= 2\pi\times 6\,$MHz). In this case, the interaction starts slightly off-resonance, but then the atoms are pushed into resonance and finally accelerated out of resonance again. Provided this transient Doppler shift is symmetric with respect to resonance, the number of emitted photons reaches its maximum. Therefore the optimum detuning depends on the probe intensity and the chosen interaction time. For the parameters given in the caption of fig.\,\ref{detune}, $N_{phot}=580$ photons per atom are expected. 

It must be noted that the acceleration could be balanced by a second, counter-propagating probe beam that provides the same excitation probability as the first one. However, there was no need to investigate this situation, simply because the required time to detect a single atom, $\Delta\tau$, is shorter than the time it takes to accelerate an atom out of resonance. 

For longer interaction times, another loss mechanism becomes significant. The small probability to excite a $^{85}$Rb atom accidentally  to  $5P_{3/2}(F'=3)$ cannot be neglected, simply because the probe laser (resonant with the $5S_{1/2}(F=3)$  to  $5P_{3/2}(F'=4)$ transition), is only detuned by $2\pi\times 121\,$MHz from the $5S_{1/2}(F=3)$ to $5P_{3/2}(F'=3)$  transition. By off-resonant excitation to $5P_{3/2}(F'=3)$ followed by spontaneous emission, the atoms can decay  to  $5S_{1/2}(F=2)$, where they remain ``in the dark'', i.e they are no longer excited by the probe laser due to the $2\pi\times 3\,$GHz hyperfine splitting between the $5S_{1/2}(F=2)$ and the $5S_{1/2}(F=3)$  states. However, even for a $2.7$-fold saturated probe transition, the   time needed to pump the atom into the other hyperfine state is about $130\,\mu$s and, hence, significantly longer than $\Delta\tau$. Therefore these losses can be neglected and no second laser is needed to pump the atoms back to the  $5P_{3/2}(F=3)$ level. 

Taking into account all these considerations,  we expect, on average,  $N_{phot}=660$ photons per atom for $I/I_{sat}=2.7$, $\Delta\tau=60\,\mu$s and an optimal probe detuning $\Delta_{opt}=0.43\,\Gamma$. Due to the $3\%$ photon detection efficiency of the collection system with the Hamamatsu PMT, 20 photons per atom can actually be counted.

\section{Signal Analysis}
\begin{figure}[tbh]
    \centering
    \includegraphics[width= \columnwidth]{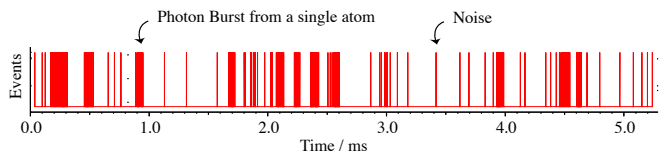}
    \caption{Stream of photons detected by the photomultiplier. A 
    burst of events indicates the presence of an atom, whereas 
    single events between the bursts belong to background noise. }
    \label{stream}
\end{figure}

To characterize the performance of the detector, we analyze the  stream of  events from the PMT. An example of a typical signal is shown in fig.\,\ref{stream}, which has been recorded with an average atom number in the observation volume far below one, i.e. the rate of atoms entering the detector, $R_{A}$, was lower than the reciprocal interaction time, $\Delta\tau^{-1}$. Several bursts of events, well separated from each other, can be distinguished. We attribute these bursts to single atoms that emit fluorescence photons at an average rate, $\langle R_{P}\rangle=N_{phot}/\Delta\tau$, while they interact with the probe beam. This gives rise to an intermittend increased photon-detection rate,  $R_{E}$, which is significantly higher than the average noise-count rate, $R_N$. From Fig.\,\ref{stream}, it is also obvious that these noise counts, which are mainly caused by scattered probe light, cannot be neglected. 

\begin{figure}[tbh]
    \centering
    \includegraphics[width= 8cm]{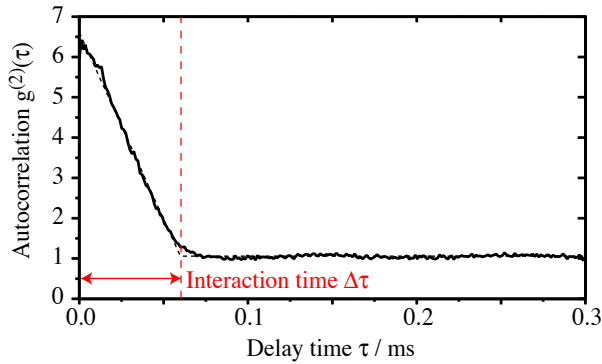}
    \caption{The solid line shows the autocorrelation of a 
    524\,ms long stream of PMT 
    events. Since every falling cloud of atoms is observed for 
    5.4\,ms, a total number of 100 experimental 
    cycles was needed to produce this data. The dashed line is a fit 
    of Eq.\,(\ref{eq:AC})  to the measured 
    data, and the vertical line indicates the interaction time 
    $\Delta\tau$.}
    \label{g2}
\end{figure}

The second order autocorrelation, $g^{(2)}(\tau)$, of the measured photon stream is well suited for the further analysis. It should only depend on the rates $R_{A}, R_{E}, R_{N}$, and $\Delta\tau$, and it is defined as
\begin{equation}
    g^{(2)}(\tau) = \frac{\langle p(t)p(t-\tau)\rangle}{\langle 
    p(t)\rangle^2},
    \label{eq:ACdef}
\end{equation}
where  $p(t)dt$ is the probability to observe a photon in the time interval $[t, t+dt]$. Photons from different atoms are not correlated, and since photons from one-and-the-same atom are  emitted only within the interaction time  $\Delta\tau$, $g^{(2)}$ must equal $one$ for $\tau>\Delta\tau$. If we assume that the event rate $R_{E}$ is constant while an atom interacts with the probe laser, a linear decrease of $g^{(2)}(\tau)$ is expected in the range $0\le\tau\le\Delta\tau$, and a  simple calculation leads to the analytical expression
\begin{equation}
    g^{(2)}(\tau) = \left\{
        \begin{array}{ccc}
        1 + \frac{R_{A}R_{E}^2 \quad (\Delta\tau -|\tau|)}{(\Delta\tau R_{A}R_{E} + R_{N})^2} 
         & \mbox{for} & |\tau| \le \Delta\tau  \\
        1 & \mbox{for} & |\tau| > \Delta\tau.
    \end{array}
    \right.
    \label{eq:AC}
\end{equation}
Note that this simple model neglects the anti-bunching of the sub-Poissonian light emitted from single atoms \cite{Kimble78}. This is well justified since this effect is only visible on a sub-$\mu$s time scale.

In the experiment, the noise rate, $R_{N}$, and the total event rate, $R=R_{N}+\Delta\tau R_{E}R_{A}$, are measured separately. Using these two values as fixed parameters, a fit of $g^{(2)}(\tau)$ from Eq.\,(\ref{eq:AC}) to the autocorrelation  of the experimental data, calculated according to Eq.\,(\ref{eq:ACdef}), allows us to determine the average number of events per atom, $\langle N\rangle=\Delta\tau R_{E}$, and the average duration of an atom's photon burst, $\Delta\tau$. Figure \ref{g2} shows the autocorrelation of a typical signal, together with the best fit to this data, yielding $g^{(2)}(0)=6.60\pm 0.03$ and $\Delta\tau=60.7\pm 0.3\,\mu$s. From these fit parameters, the measured noise rate, $R_{N}=9.4\pm 0.13\,$kHz, and the total number of 22232 photon events in $524\,$ms (i.e. $R=42.4\pm 0.29\,$kHz), we know that a single atom causes an average number of $\langle N\rangle=20.4\pm 0.3$ events. This is in good agreement with the theoretical expectation, $N=20$.

\begin{figure*}[tbp]
    \centering
    \includegraphics[width= 14cm]{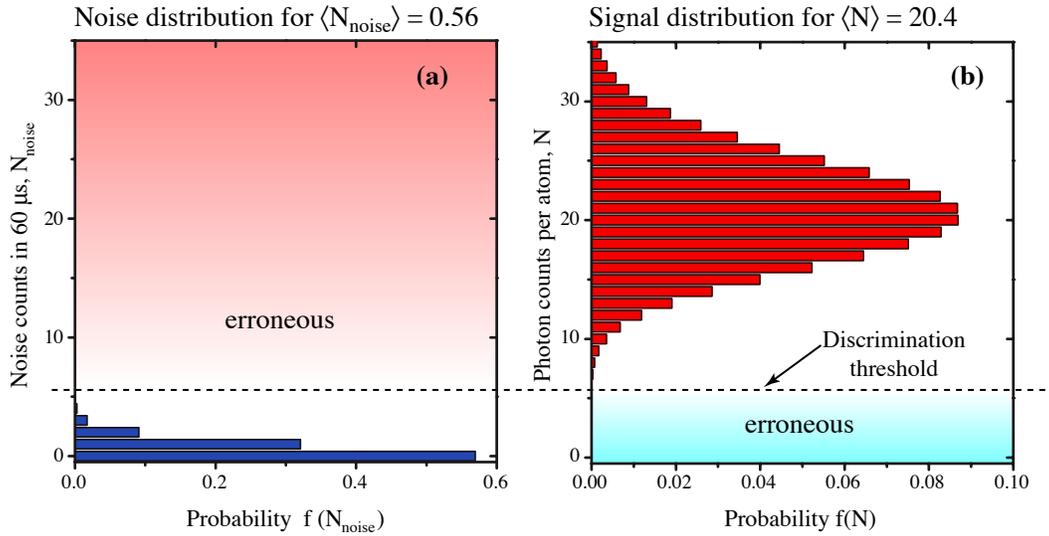}
    \caption{Assumed Poissonian distribution of the number of noise \textbf{(a)} and 
    signal \textbf{(b)}  events for a measurement interval of $\Delta\tau=60\,\mu$s. The 
    discrimination level is indicated and the corresponding erroneous regions are 
    shown as shaded areas.}
    \label{Poisson}
\end{figure*}

For the time-resolved detection of single atoms, it is essential to discriminate their signal from the background noise. Unfortunately, this noise is mainly caused by light scattered  from the probe beam. Hence, it has the same wavelength as the fluorescence photons and  cannot be eliminated by interference filters. However, spatial filtering reduces the noise rate to $R_{N}=9.4\,$kHz for a 2.7-fold saturated probe laser with an intensity of 10\,mW/cm$^2$, a height of 0.2\,mm and a width of 0.7\,mm. Under these circumstances, the event rate rises to $R_{E}\approx 340\,$kHz when an atom is in the observed volume, and the signal-to-background-noise ratio is $\approx 36$. In the interesting time interval of  $\Delta\tau=60\,\mu$s, an average number of noise events, $N_{noise}=0.56$, is expected, while the average number of events for a single atom is  $\langle N\rangle = 20.4$. Figure\,\ref{Poisson} shows Poissonian probability distributions of the noise and the signal events for a $60\,\mu$s long discrimination interval. For these distributions, the optimum threshold to distinguish atoms from noise is  $N\ge 6$. The possible errors can be estimated from  Poissonian statistics: On the one hand, the probability to attribute more than five noise events in an arbitrarily  chosen $60\,\mu$s long interval to be due to an atom is as low as 0.0027\,\%, i.e. with a probability of 21\,\% the noise gives rise to a single ``fake atom'' during the total measurement time of 524\,ms. On the other hand, the probability to miss an atom because it does not emit enough photons is only 0.0034\,\%, i.e. 0.03 atoms from the 945 atoms counted in 524\,ms are missing, which is negligible.

\begin{figure*}[tbp]
    \centering
    \includegraphics[width= 14cm]{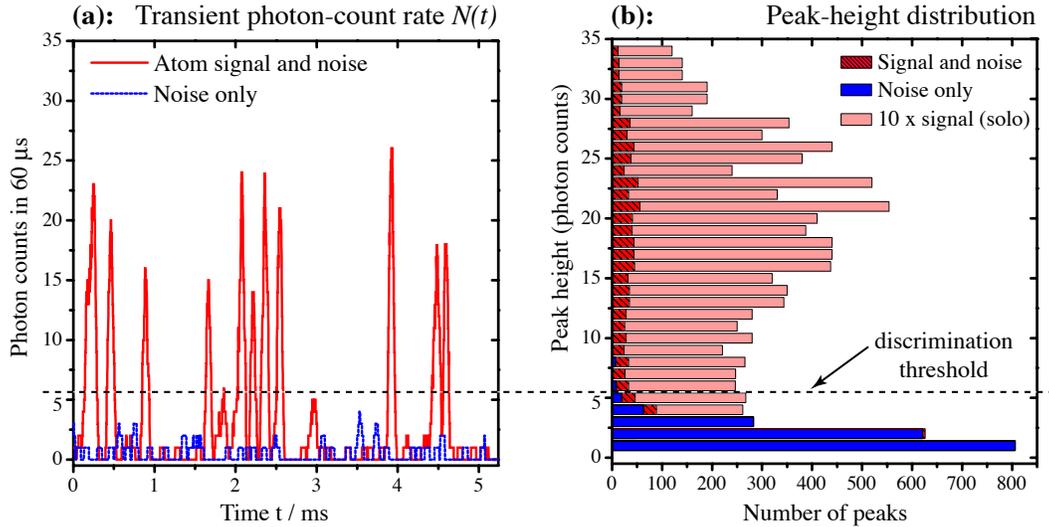}
    \caption{Peak height analysis of the measured signal. \textbf{(a):} 
    Photon-detection rate (number of counts in 60\,$\mu$s) as a 
    function of time. The evaluation makes use of an integration 
    interval which is sliding smoothly along the stream of measured 
    events, not of a fixed grid with $60\,\mu$s spacing. 
    Hence the borders of the corresponding interval
    coincide with the atom's arrival and departure times in each peak. For
    comparison, the same  analysis is also shown for the noise level (without atoms).
    \textbf{(b):} Measured distribution of the peak count rates. Since 
    only peaks are considered, there is no signal with \textit{zero} 
    counts.}
    \label{Poisson2}
\end{figure*}

Based on these considerations, we determine the arrival times of individual atoms from the registered photon stream with the help of the transient photon-count rate, $\tilde{N}(t)$, which is defined as the number of photons counted in the interval $[t, t+\Delta\tau]$. For every time $t$ that coincides with the arrival time of an atom, all its photons fall into this interval, and $\tilde{N}(t)$ reaches a peak value that coincides with the number of photons emitted from the atom. Figure\,\ref{Poisson2}(a) shows the transient photon-count rate  that belongs to the data from Fig.\,\ref{stream}. We easily identify several peaks above threshold which indicate the arrival times of single atoms. For comparison, a trace without atoms is also shown, and obviously, the signal is always below the  threshold for atom detection. Figure\,\ref{Poisson2}(b) shows the distribution of the peak heights for two photon streams, one with atoms, and the other one without. Some deviations from the ideal situation are evident. The numbers of detected photons per atom, $N$, scatter over a  much larger range than expected from Poissonian statistics. We attribute this large variation to the different Clebsch-Gordan coefficients of the randomly populated magnetic sublevels, which give rise to a large variation of the saturation intensities of the involved transitions, i.e. $I_{sat}\approx 2.9 \ldots 6.6\,$mW/cm$^2$ for $\pi$-polarized light, and $I_{sat}\approx 1.6 \ldots 46.4\,$mW/cm$^2$ for $\sigma$-polarized light. However, it is evident that most atoms can be well distinguished from noise, which itself  is rarely identified as a ``fake atom''.

We expect that this simple, fast and state-selective way of detecting single atoms will be very useful in cavity-based quantum information processing. It allows one to probe the presence and the quantum state of individual atoms with high accuracy. A slight drawback is the momentum kick the atoms experience during detection, which might restrict application of our scheme to post selection.

\section{Summary}
We have shown that laser-induced fluorescence from a saturated and (nearly) closed two-level atomic transition allows one to monitor single, freely moving atoms with a time resolution of $60\,\mu$s. To collect the emitted photons, a mirror setup  covering a very large solid angle is used, which focuses the photons onto a photomultiplier tube. A careful analysis of the measured data based on a second-order autocorrelation shows that, on average, 20.4 photons per atom are detected. In case of a Poissonian distribution of noise and signal events, this signifies that a single atom is detected with 99.9966\,$\%$ certainty, while noise counts exceed threshold with a negligible probability of 0.0027\,\%. 
Even the non-Poissonian photon distribution, due to  the manyfold of magnetic sublevels  in our experiment, barely affects the single-atom detection efficiency.
In particular for quantum information science, this highly reliable, fast and state-selective detection of single atoms will be a very useful tool. 

\section*{Acknowledgments}
This work was partially supported by the  Deutsche Forschungsgemeinschaft (SPP 1078, SFB 631, research unit 635) and the European Union in the IST (QGATES, SCALA) and IHP (CONQUEST) programs.

We dedicate this article to Bruce W. Shore at the occasion of his 70$^{th}$ birthday.

\end{document}